\documentstyle[prl,aps]{revtex}
\begin{document}
\draft

\twocolumn[\hsize\textwidth\columnwidth\hsize\csname@twocolumnfalse\endcsname

\title{Dynamic transition in vortex flow in strongly disordered 
Josephson junction arrays and superconducting thin films}

\author{Daniel Dom\'{\i}nguez}
\address{Centro At\'{o}mico Bariloche, 8400 S. C. de Bariloche,
Rio Negro, Argentina}

\date{\today}
\maketitle
\begin{abstract}
We study the dynamics of vortices in strongly disordered $d=2$ 
Josephson junction  arrays and superconducting films driven by a current.
We find a dynamic phase transition in vortex flow at a current $I_p>I_c$.
Below $I_p$ there is plastic flow  characterized by 
an average-velocity correlation length scale $\xi_v$ in 
the direction of motion, which diverges
when approaching $I_p$. Above $I_p$ we find
a moving vortex phase with homogeneous flow and short range 
smectic order. A finite size analysis
shows that this phase becomes asymptotically a liquid for large length scales.
\end{abstract}


]                

\narrowtext

The study of non-equilibrium steady
states of driven many-degrees-of-freedom systems with quenched
disorder are of importance in many condensed matter systems
\cite{SB,KV,BF,plastic,filam,olson,GLD,bmr,scheidl,bhatta,aharon,yaron,pardo,%
moon,ryu,spencer,dgb,jjaexp,perco,nos,dyna}.
Examples of this problem are the dynamics of vortices
in type II superconductors \cite{SB,KV} and charge density waves
\cite{BF}.
For  low driving
forces the dynamics is dominated by disorder leading to
a plastic flow regime
\cite{SB,KV,plastic,filam,olson}. 
On the other hand, for very large driving forces  the randomness 
should  be less important and all the internal
degrees of freedom will move more or less coherently as a whole \cite{SB,KV}. 
Recently, Koshelev and Vinokur \cite{KV} have 
proposed that  there is an ordered moving vortex phase.
However, Giamarchi and Le Doussal \cite{GLD} have shown that
some modes of static disorder are still
present in the moving system, leading to a moving Bragg glass (MBG) phase
\cite{GLD}.
In turn, Balents, Marchetti and Radzihovsky \cite{bmr}
 have argued that the driven
state is a moving smectic (MS), consisting on liquid channels
with transverse periodic order. Experimentally, studies of current-voltage
characteristics \cite{bhatta,aharon}
and neutron-scattering experiments \cite{yaron} have found 
a reordering of the vortex structure when increasing the current 
bias. Recently, Pardo {\it et al.} \cite{pardo}
have found in decoration experiments that for low magnetic 
fields there is a MS vortex structure  while for high fields there is a MBG.
Numerical simulation studies have also found an ordering of the vortex system
for high currents \cite{SB,KV,moon,ryu,spencer,dgb}. 
Moon {\it et al} \cite{moon} have found a MS phase
in $d=2$ molecular dynamics simulations  with a short-range interaction
potential,  while in \cite{dgb} a MBG phase was found in a driven XY model
simulation for $d=3$ and large magnetic fields.
Therefore, questions such us the existence and nature of a  moving
phase and  which effects of the disorder remain once the vortices 
are in motion are currently under discussion.

Most of the experimental systems mentioned above have the difficulty
that there is no control of the nature and amount of disorder.
Therefore the study of disordered  Josephson junction arrays (JJA) 
becomes particularly promising here, since they can be specifically
fabricated with controlled randomness \cite{jjaexp}. 
For example, questions such as the dynamics
of JJA with percolation disorder \cite{jjaexp,perco} 
and the plastic flow of vortices have been studied recently \cite{nos}. 
Also, there is an intrinsic interest
in the non-linear dynamics of JJA and their non-equilibrium properties,
see for example \cite{dyna}. 
In this paper we will show that there is a dynamic transition
in driven JJA  with strong positional disorder.

The current flowing in  the junction between two superconducting islands 
in a JJA is 
modeled as the sum of the Josephson supercurrent and the normal current
\cite{perco,nos,dyna}:
\begin{equation}
I_{\mu}({\bf n})= I^0_{\mu}({\bf n})\sin\theta_{\mu}({\bf n}) + 
                  \frac{\Phi_0}{2\pi c R_N} 
		  \frac{\partial \theta_{\mu}({\bf n})}{\partial t}
\end{equation}
where $I^0_{\mu}({\bf n})$ is the critical current of the junction between
the sites ${\bf n}$ and ${\bf n}+{\bf \mu}$ in a square lattice 
[${\bf n}=(n_x,n_y)$, ${\bf \mu}={\bf \hat x}, {\bf \hat y}$], $R_N$ is the normal state resistance
and $\theta_{\mu}({\bf n})=\theta({\bf n}+{\bf \mu})-\theta({\bf
n})-A_{\mu}({\bf n})=\Delta_\mu\theta({\bf n})-A_{\mu}({\bf n})$ is the
gauge invariant phase difference with $A_{\mu}({\bf n})=\frac{2\pi}{\Phi_0}
\int_{{\bf n}a}^{({\bf n}+{\bf\mu})a}{\bf A}\cdot d{\bf l}$. 
In the presence of an external magnetic field $H$ we have
$\Delta_{\mu}\times A_{\mu}({\bf n})= A_x({\bf n})-A_x({\bf n}+{\bf y})+ 
A_y({\bf n}+{\bf x})-A_y({\bf n})=2\pi f$, $f=H a^2/\Phi_0$ and 
$a$ is the array lattice spacing.
Here we  consider a distribution of critical currents $I^0_{\mu}({\bf n})=
I_0\delta_{\mu}({\bf n})=I_0[1+\delta(RAN-1/2)]$ with $RAN$ a 
random uniform number in $[0,1]$.
We take periodic boundary  conditions (p.b.c) in both directions in the presence
of an external current $I_{ext}$ in the $y$-direction in arrays with 
$L\times L$ junctions.
The vector potential is taken as
$A_{\mu}({\bf n},t)=A_{\mu}^0({\bf n})-\alpha_{\mu}(t)$ where 
in the Landau gauge $A^0_x({\bf n})=-2\pi f n_y$, $A^0_y({\bf n})=0$
and $\alpha_{\mu}(t)$ will allow for total voltage fluctuations. 
With this gauge the p.b.c. for the phases are: 
$\theta(n_x+L,n_y)=\theta(n_x,n_y)$ and
$\theta(n_x,n_y+L)=\theta(n_x,n_y)-2\pi f Ln_x$. 
The condition  of a current flowing in the $y$- direction:
$\sum_{\bf n} I_{\mu}({\bf n})=I_{ext}L^2\delta_{\mu,y}$
determines the dynamics of $\alpha_\mu(t)$ \cite{vincent}.
After considering conservation of current, 
$\Delta_\mu\cdot I_{\mu}({\bf n})=\sum_{\mu} I_{\mu}({\bf n})-
I_{\mu}({\bf n}-{\bf \mu})=0$, we obtain the equations:
\begin{eqnarray}
\Delta_{\mu}^2\frac{\partial\theta({\bf n})}{\partial t}&=&-\Delta_{\mu}\cdot
S_{\mu}({\bf n})\\
\frac{\partial\alpha_{\mu}}{\partial t}&=&I_{ext}\delta_{\mu,y}
-\frac{1}{L^2}\sum_{\bf n} S_{\mu}({\bf n})
\end{eqnarray}
where $S_{\mu}({\bf n})=\delta_{\mu}({\bf n})\sin[\Delta_\mu\theta({\bf n})-A_{\mu}^0({\bf n})-
\alpha_{\mu}]$, we have normalized currents by $I_0$ and time
by $\tau_J=2\pi cR_{N}I_0/\Phi_0$, and we have defined the discrete Laplacian
$\Delta^2_\mu\theta({\bf n})=\theta({\bf n}+{\bf\hat x})
+\theta({\bf n}-{\bf\hat x})+\theta({\bf n}+{\bf\hat y})
+\theta({\bf n}-{\bf\hat y})-4\theta({\bf n})$.
These same equations represent the dynamics of a superconducting thin film
(STF) after discretization  of a time-dependent London model.
One starts with the current density as the sum of supercurrent and normal
current:
\begin{eqnarray}
{\bf J}&=&{\bf J}_S+{\bf J}_N\nonumber\\
{\bf J}&=&\frac{ie\hbar}{m^*}
[\Psi^*{\bf D}\Psi-({\bf D}\Psi)^*\Psi]+
\frac{\sigma\Phi_0}{2\pi c}\frac{\partial}{\partial
t}(\nabla\theta-\frac{2\pi}{\Phi_0}{\bf A})
\end{eqnarray}
with ${\bf D}=\nabla+i\frac{2\pi}{\Phi_0}{\bf A}$
and $\Psi({\bf r})=|\Psi({\bf r})|\exp[i\theta({\bf r})]$.
One takes the discretization ${\bf r}=(n_x\xi,n_y\xi)=\xi{\bf n}$
with the rule $D_\mu \Psi(\bf {r})\rightarrow\frac{1}{\xi}\{\Psi({\bf n}+
{\bf\mu})-\exp[-i\frac{2\pi}{\Phi_0}A_\mu({\bf n})]\Psi({\bf n})\}$.
After considering conservation of current, and assuming $|\Psi({\bf n})|$ 
is quenched and depends only on disorder, one obtains the same equations
as in (2-3). Now $I_\mu({\bf n})$ has to be interpreted as current density
normalized by $J_0=2e\hbar |\Psi_0|^2/m\xi=\Phi_0/(8\pi^2\lambda^2\xi)$,
time normalized by $\tau=c/(4\pi\sigma\lambda^2)$, 
$\delta_\mu({\bf n})=|\Psi({\bf n}+{\bf\mu})||\Psi({\bf n})|/|\Psi_0|^2$,
and the field density is $f=H\xi^2/\Phi_0=H/2\pi H_{c2}$.
The $T=0$ dynamical equations (2-3) are solved with a second order Runge-Kutta
algorithm with time step $\Delta t=0.05\tau_J$ and integration time $1000\tau_J$
after a transient of $500\tau_J$. The discrete Laplacian is inverted with
a fast Fourier + tridiagonalization algorithm as in \cite{nos}.

We consider here very strong disorder with $\delta=0.5$. 
For STF, this corresponds to 
a extremely dense distribution of pinning sites with a pinning potential
with a $50\%$ fluctuation in amplitude in the length scale of $\xi$,
while for JJA it corresponds to a $50\%$ fluctuation in the critical 
currents. 
The ground state $I_{ext}=0$ vortex configuration
is a vortex glass with no structure in the structure factor
(no Bragg peaks).
We study a magnetic field of  $f=1/25$
and system sizes of $L=50,100,150,200$. We calculate
the time average of the total voltage  $V=\langle v(t)\rangle=
\langle d\alpha_y(t)/dt\rangle$  (normalized by $R_NI_0$)
as a function of $I$ as shown in Fig.1(a) for one sample of size $L=200$.
The error bars of $V$  [obtained from the statistics of the time
averaging of $v(t)$] are smaller 
than the symbol size in Fig.~1(a) (error in $V$ $\sim10^{-5}-10^{-4}$).  
This suggests that the sample
size is large enough to be self-averaging.
The initial condition is a thermally quenched vortex configuration
at $I=0$. From this state the current $I$ is slowly 
increased in steps of $\Delta I=0.01$ taking
as initial condition the last phase configuration of the previous current.
We obtain similar results by slowly decreasing decreasing the current
from a random vortex configuration at $I=0.8$. 
Above a critical current of $I_c=0.105$ (in units of $I_0$) 
there is a non-linear onset
of voltage with plastic flow of vortices.
We study the time-averaged voltage in the
bonds parallel to the direction of the current drive: 
$v_y({\bf n})=\langle d\theta_y({\bf n},t)/dt\rangle$, which is proportional
to the average vortex speed in the direction of the Lorentz force.
We see in Fig.2(a) that near $I_c$ vortex flow is very inhomogeneous,
as typical for the plastic flow regime, showing channels of flow for low
currents. For increasing drives the flow becomes more homogeneous as
shown in Fig.2(b). We characterize the inhomogeinity of the 
flow  with the correlation function for voltages along
the direction of motion: 
$C_v(x)=\frac{1}{L^2}\sum_{{\bf n}}v_y({\bf n})v_y({\bf n}+x\hat{\bf x})-
\langle v_y\rangle^2$.
We see in Fig.2(c) that the voltage correlation increases for increasing
values of $I$. There is a characteristic correlation length $\xi_v$ 
defined by $C_v(x)\approx C_v(0)\exp(-x/\xi_v)$. This voltage correlation 
length was proposed by Bhattacharya and Higgins \cite{bhatta} 
as a characteristic length
scale for the dynamics of plastic flow. 
We see in Fig.1(b) that $\xi_v$ increases
with current $I$ and it diverges ($\xi_v\gg L$) at a current $I_p(L)$.
We have also analyzed the voltage correlation function along $y$, $C_v(y)$.
In this case, there is always a fast decay of $C_v(y)$ for any bias current,
therefore above $I_p$ the voltage distribution $V_y({\bf n})$ 
becomes homogeneous along the $x$ direction but it always has fluctuations
along the $y$ direction. This reflects the fact that for large drives
the effect of the disorder potential becomes negligible only along the
direction of the Lorentz force, but along the transverse direction is still
important \cite{GLD,bmr}.
The size dependence of $I_p(L)$ is shown in the inset of Fig.1(b).
We see that in the limit of large $L$, $I_p$ tends to a finite value
of $I_p \approx 0.31$ in this case.
This shows that in the thermodynamic limit there is a {\it dynamic
phase transition}.

Above $I_p$, we 
study the ordering of the moving vortex structure. In order to follow 
the vortex  positions directly, we obtain the vorticity
at the plaquette ${\bf \tilde n}$ (associated to the site ${\bf n}$)
as $b({\bf \tilde n})=-\Delta_\mu\times{\rm nint}[\theta_\mu({\bf n})/2\pi]$
with ${\rm nint}[x]$ the nearest integer of $x$. We calculate the time-averaged
vortex structure factor as $S({\bf k})=|\langle\frac{1}{L^2}\sum_{\bf \tilde n}
b({\bf \tilde n})\exp(i{\bf k}\cdot{\bf \tilde n})|^2$. For currents $I<I_p$,
$S({\bf k})$ has only the density peak $S({\bf k}=0)=f^2$ and an 
isotropic ring-like
structure as expected for plastic flow \cite{moon,ryu,spencer}. 
On the other hand, for $I>I_p$ 
there are well-defined peaks  
in $S({\bf k})$ as shown in the surface plot of Fig.3(a) for $L=150$. 
We see that there are two strong peaks  in the $k_y$ direction, 
at vectors ${\bf K}=\pm Q_s{\bf \hat y}$, 
consistent with smectic ordering in the direction transverse 
to motion \cite{bmr}. 
($Q_s=2\pi/a_s$ with $a_s\approx4.5\approx(\sqrt{3}/2f)^{1/2}$
the row spacing in a triangular lattice).
There are also small
satellite peaks in the other ${\bf K}$ directions. 
The position of the peaks is better seen in the two dimensional 
gray scale plot of Fig.~3(b). We see that the strongest spots besides
${\bf K}=\pm Q_s{\bf \hat y}$ are
at the reciprocal space ${\bf K}$-vectors corresponding 
to a triangular lattice, suggesting the presence of some orientational order.
\cite{teitel}
In Fig.~3(c) we show a finite size analysis of $S(Q_s)\sim L^{-\nu_s}$.
For a moving smectic with quasi-long range order one expects $\nu_s<2$,
while $\nu_s=2$ is the value for a liquid.
We obtain $\eta_s=1.96\pm0.06\approx2$.
Therefore in the thermodynamic
limit vortex flow is always liquid-like in nature for sufficiently large
length scales, and smectic ordering is only a short-range phenomenon in 
$d=2$, at least for strong disorder. 
 This is consistent with RG calculations which
find that a moving smectic phase might be unstable 
in $d=2$ \cite{bmr}. 
In Fig.1(c) we show the parameter of short-range 
smectic order  $S(Q_s)$ as a function of
$I$ for a lattice size $L=200$. 
We find that for large drives $S(Q_s)$ is
nearly current independent, it decreases when decreasing $I$ and vanishes
when approaching $I_p$ from above. This convalidates the
result that there is a dynamic transition at $I_p$ where there is an
onset of  anisotropic short range order.

The  ``flux flow noise'' has been studied in  many 
current-voltage measurements \cite{yeh,noise}. 
Here we analyze the voltage noise response with the 
power spectrum $P(\nu)=|\frac{1}{T}\int_0^Tdt v(t)\exp(i2\pi\nu t)|^2$.
We find that in the plastic flow regime, very near to $I_c$, the power
spectrum shows $1/\nu$ noise, see Fig.4(a).
On the other hand, for large drives $I>I_p$, the power spectrum tends
to a frequency independent value for $\nu\rightarrow 0$ as shown in Fig.4(b).
These two types of power spectra: for $I<I_p$ and
$I>I_p$ have also been found in the experiments \cite{yeh,noise}.

In conclusion, we find a dynamic phase transition 
at a current $I_p$ above the critical current $I_c$. This
transition is between two different types
of liquids which differ in their {\it dynamics}  
and the spatio-temporal nature of the flow. 
Below $I_p$ there is a plastic flow regime (``turbulent'')
with an isotropic liquid structure.
The inhomogeneity of the flow is
characterized by a finite average-voltage correlation length \cite{bhatta}. 
Above $I_p$ there is a  regime with homogeneous flow
in the direction of motion (``laminar'') and with a 
structure with anisotropic short range order.
Finite size analysis shows that 
moving smectic order is only
a short-range phenomenon in $d=2$ \cite{bmr}. 
These moving vortex phases could be observable
experimentally in Josephson junction arrays using vortex-imaging techniques.

D. D. acknowledges Fundaci\'{o}n
Antorchas  and Conicet (Argentina) for financial support.

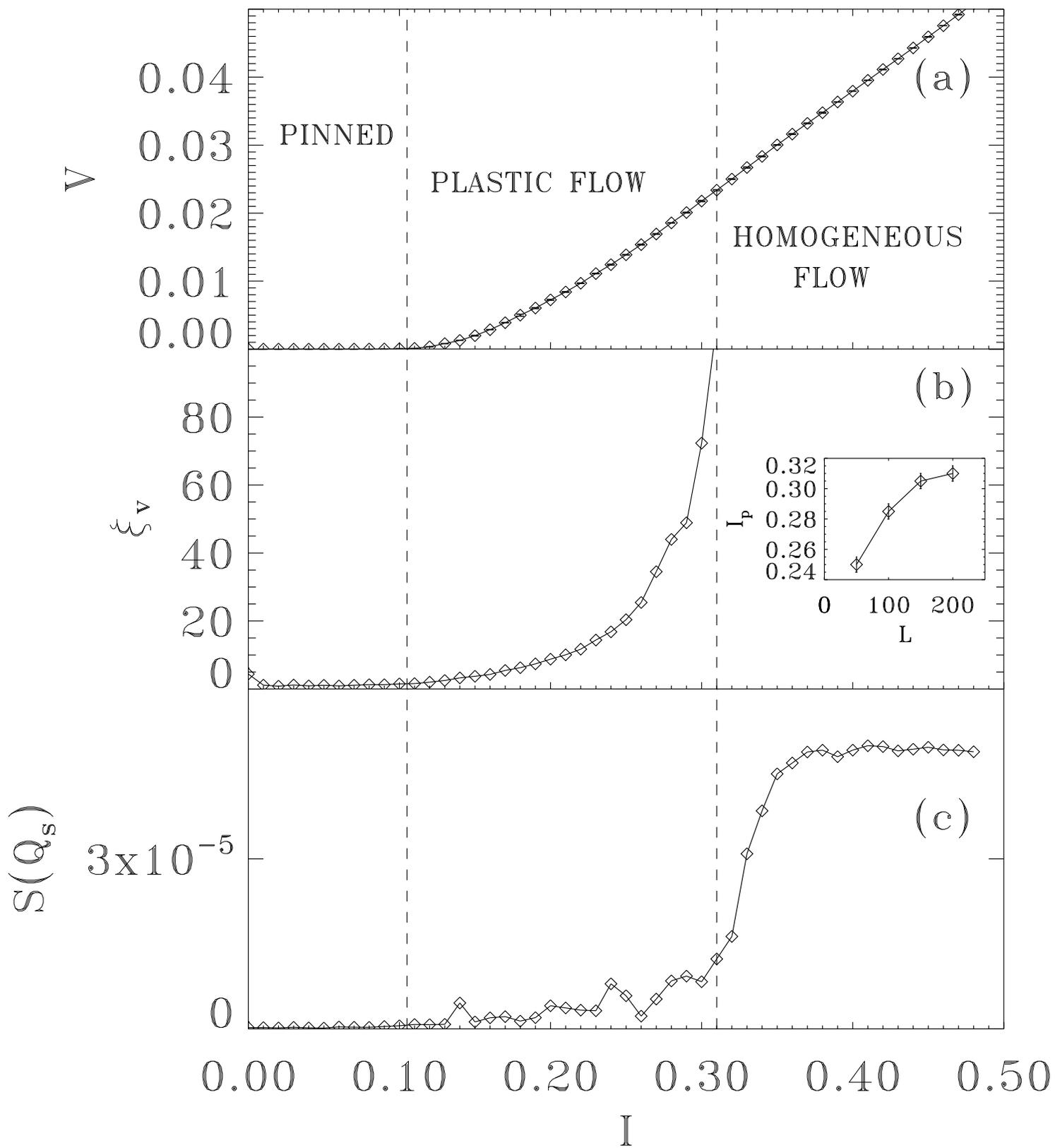
\begin{figure}

\caption{(a) Voltage  vs. current for a $200\times200$ Josephson
junction array (or a $200\xi \times 200\xi$ superconducting
thin film, STF) with field
$f=1/25$ ($B/H_{c2}=2\pi/25$ for a STF) 
and disorder $\delta=0.5$.
(b) $\xi_v$: average voltage correlation length  along the direction
of the Lorentz force. Inset: finite size dependence of $I_p$.
(c) $S(Q_s)$: Intensity of the Bragg peak for smectic order.}
\end{figure}

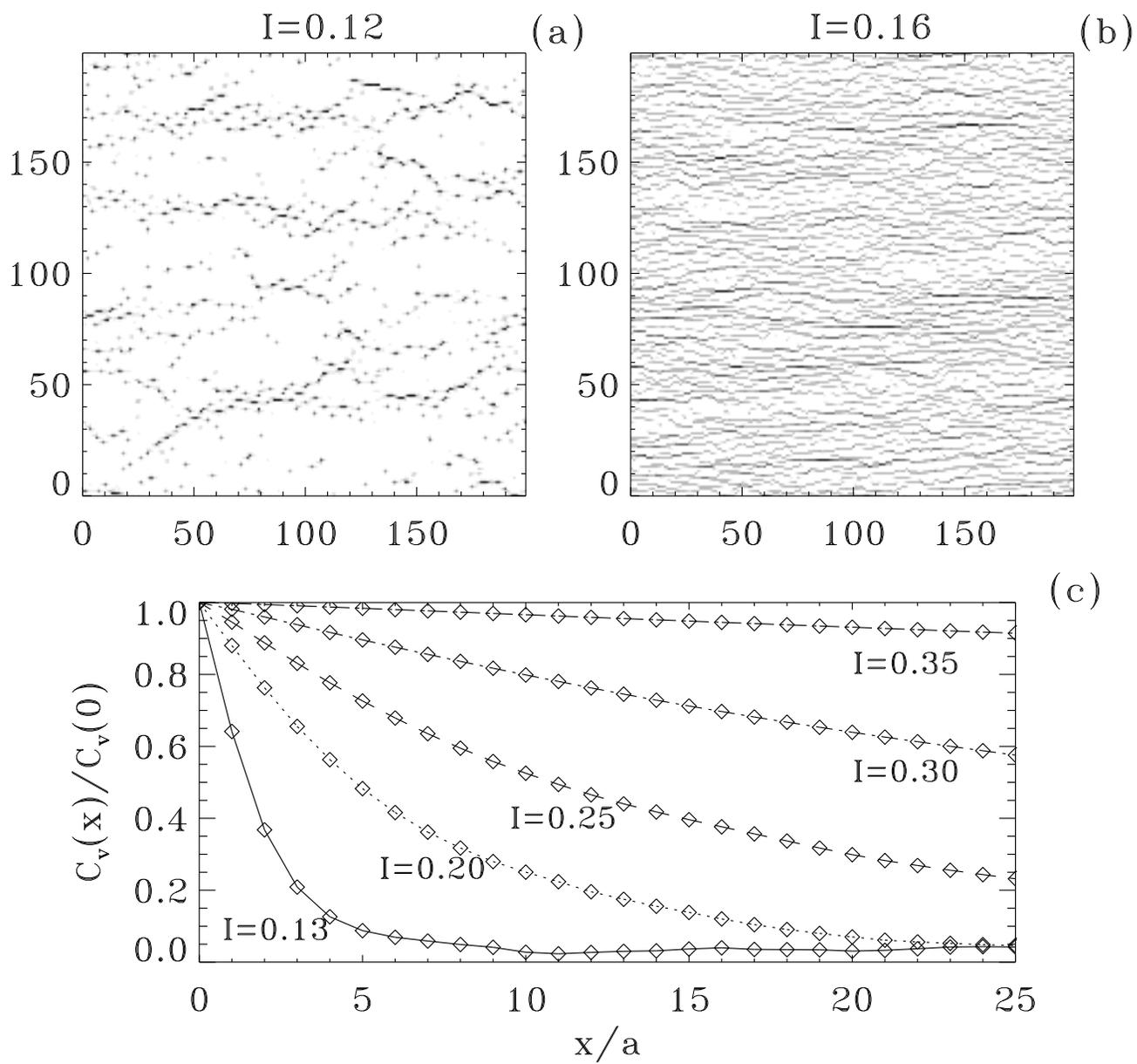
\begin{figure}
\caption{Time-averaged voltage distribution $v_y({\bf n})$
for low currents in the plastic flow regime. The gray scale is proportional to
the voltage intensity. (a) $I=0.12I_0\approx 1.2 I_c$.
(b)$I=0.16I_0\approx 1.6 I_c$
(c) $C_v(x)$: correlation function of the average voltage along
the direction of vortex motion.}
\end{figure}

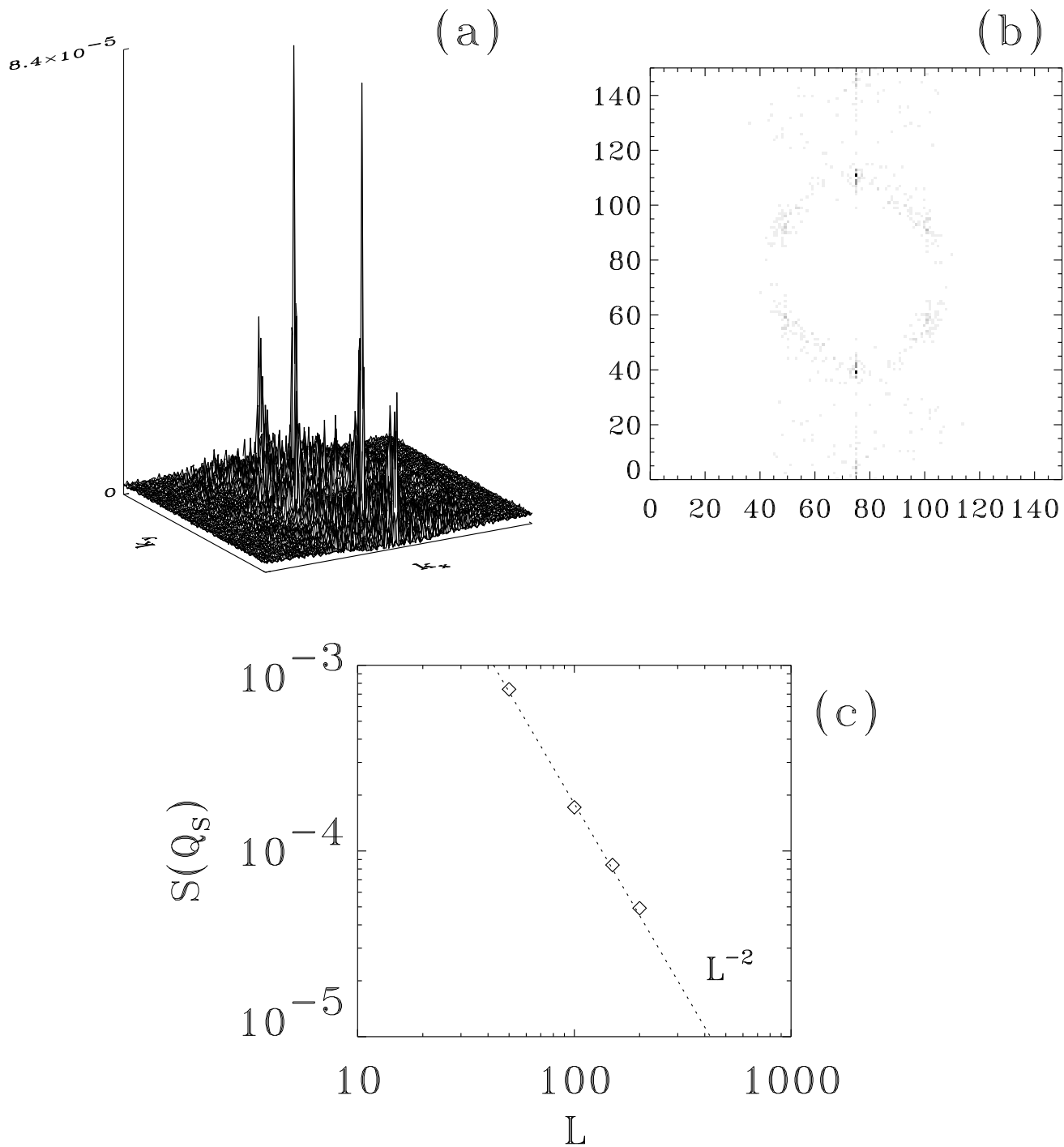
\begin{figure}
\caption{(a) Surface intensity plot of the structure factor $S({\bf k})$ of
the moving vortex system for a current $I=0.40I_0\approx 4 I_c$ and $L=150$.
We set the central peak $S(0)=0$ for clarity. 
(b) Gray scale intensity  of $S({\bf k})$.
(c) Finite size analysis of the smectic order $S(Q_s)$.}
\end{figure}

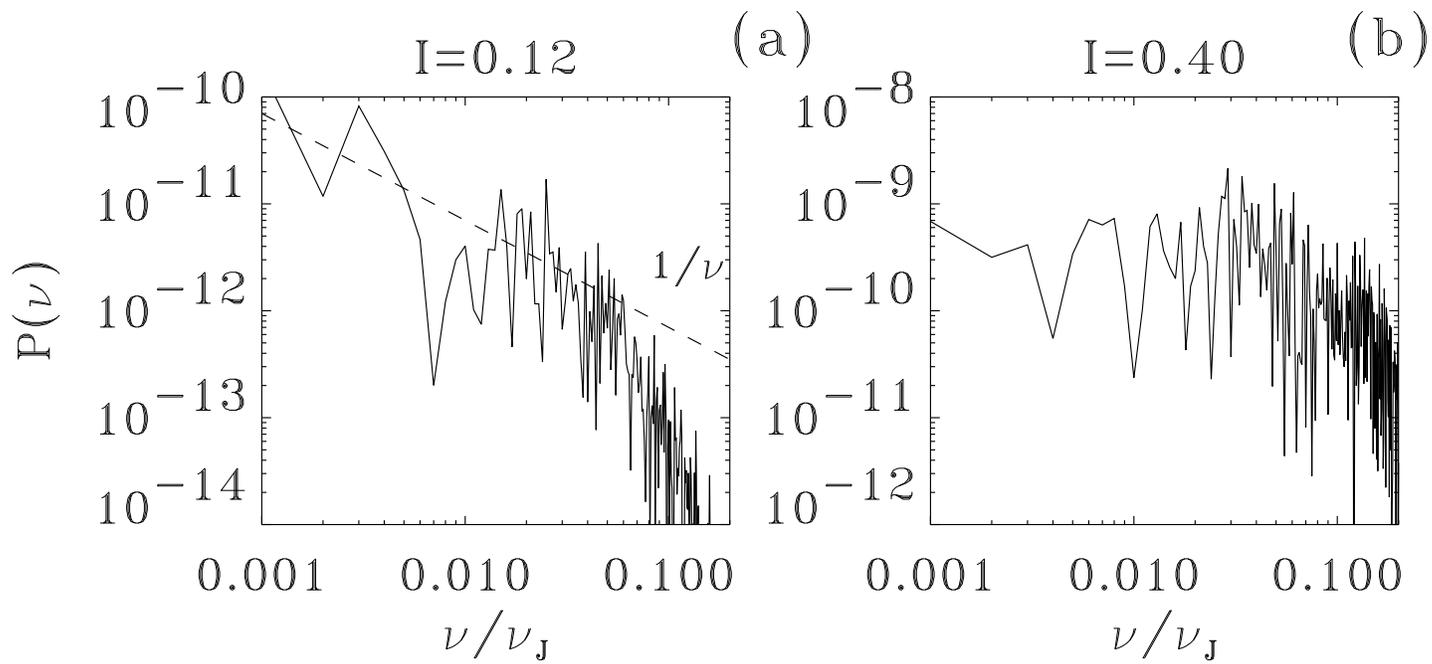
\begin{figure}
\caption{Voltage power spectrum $P(\nu)$, with
frequencies normalized by $\nu_J=\frac{\Phi_0}{2\pi cR_{N}I_0}$
(for a STF, $\nu_J=4\pi\sigma\lambda^2/c$). 
(a) $I=0.14I_0\approx 1.4 I_c$.
(b)$I=0.40I_0\approx 4.0 I_c$.}
\end{figure}

\end{document}